\documentclass[3p,times]{elsarticle}

\usepackage{ecrc}

\volume{00}

\firstpage{1}

\journalname{Nuclear Physics A}

\runauth{O.~Kaczmarek}

\jid{nupha}

\usepackage{amssymb}

\biboptions{}

\usepackage[figuresright]{rotating}

\def\lsim{\raise0.3ex\hbox{$<$\kern-0.75em\raise-1.1ex\hbox{$\sim$}}}
\def\gsim{\raise0.3ex\hbox{$>$\kern-0.75em\raise-1.1ex\hbox{$\sim$}}}

\begin{document}

\begin{frontmatter}



\dochead{}

\title{Recent Developments in Lattice Studies for Quarkonia}


\author{O.~Kaczmarek}

\address{Fakult\"at f\"ur Physik, Universit\"at Bielefeld, D-33615 Bielefeld, Germany}

\ead{email okacz@physik.uni-bielefeld.de}

\begin{abstract}
After discussing results of dilepton rates and electrical conductivity obtained
from continuum extrapolated results of light quark correlation functions in
quenched QCD
I will give a review on recent developments in lattice QCD studies for quarkonia in
the quark gluon plasma. 
Recent progress in the extraction
of spectral properties from lattice QCD calculations of hadronic correlation
functions will be discussed.
Besides medium modifications of bound states and their dissociation in the
plasma I will focus on transport coefficients, like heavy quark diffusion
constants extracted from different correlation functions on the lattice.
Present limitations and future perspectives for studies of quarkonia and
related transport coefficients on the lattice will be discussed.
See also \cite{Muller:2012hr,Gale:2012xq,Rothkopf:2012et,Ding:2012ar} for recent overviews
and related discussions of those topics.
\end{abstract}


\end{frontmatter}

\section{Dilepton rates and electrical conductivity}
\label{sec:dilepton}

A detailed knowledge of the spectral function of mesonic states at high
temperatures is of fundamental importance for the understanding of the
properties of the quark gluon plasma as well as for the analysis of this hot
and dense medium in heavy ion collision experiments.
In lattice QCD calculations the spectral function itself is not directly
calculable but can be derived from hadronic correlation functions.
In this section we will focus on the vector channel
in the light quark sector before discussing heavy quark results in the
remaining sections.
The current-current correlation functions can be represented in terms of
an integral over spectral functions, $\rho_{\mu\nu}(\omega,\vec{p},T)$.
We denote by $\rho_{ii}$ the sum over the three
space-space components of the vector spectral function and also introduce the
vector spectral function $\rho_V \equiv \rho_{00}+\rho_{ii}$. With this 
we obtain the corresponding correlation functions,
\begin{figure}[tphb]
\centering
\includegraphics[scale=0.7]{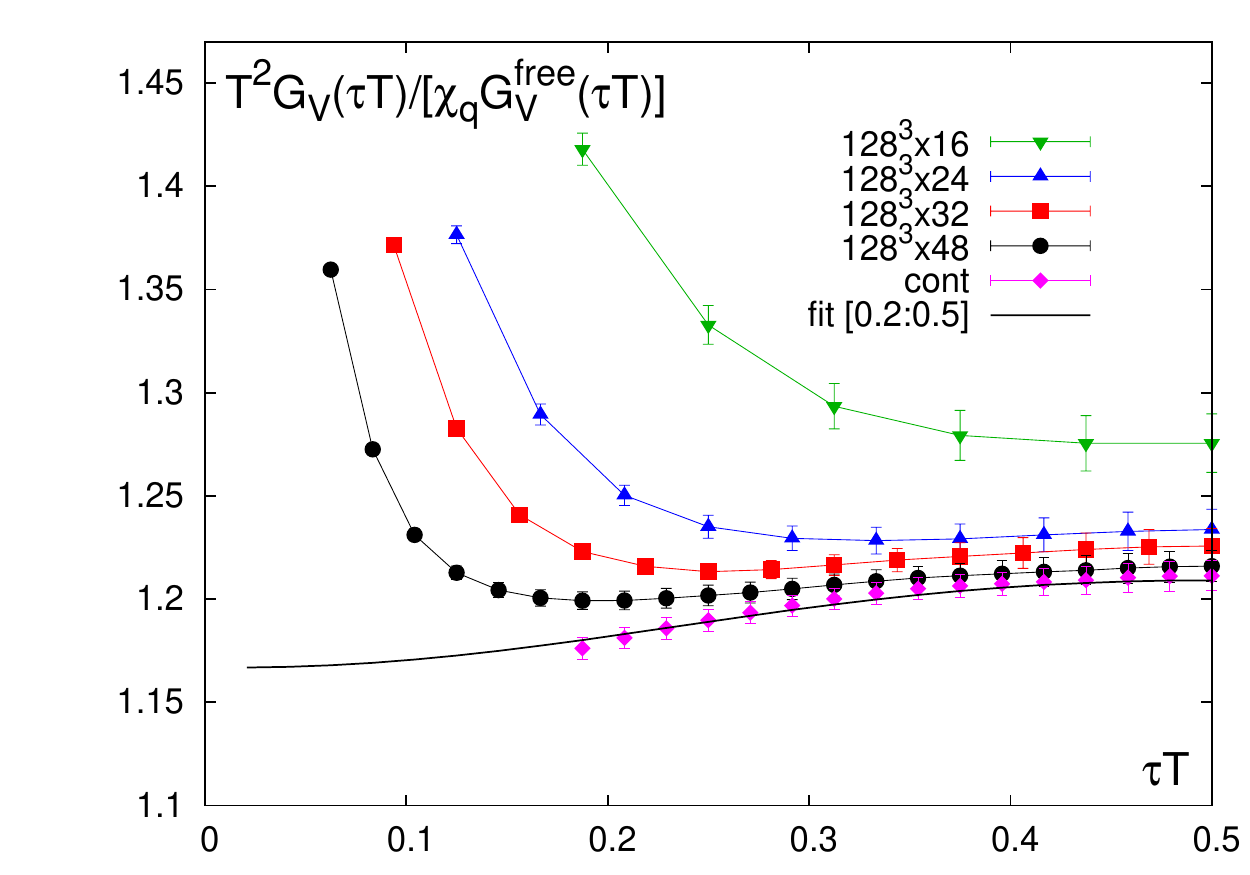}
\caption{Lattice QCD results together with the continuum extrapolation of the
  vector correlation function at $1.46~T_c$ from \cite{Ding:2010ga}.
}
\label{fig:GVcont}
\end{figure}
\begin{equation}
G_{H}(\tau,\vec{p},T) =  \int_{0}^{\infty} \frac{{\rm d} \omega}{2\pi}\;
\rho_{H} (\omega,\vec{p},T)\;
{{\rm cosh}(\omega (\tau - 1/2T)) \over {\rm sinh} (\omega /2T)}
\quad , \quad H=00,\ ii, \ V \ .
\label{speccora}
\end{equation}
The vector spectral function 
is directly related to the thermal production rate of dilepton pairs
with squared invariant mass $\omega^2 - \vec p^2$, 
\begin{equation}
{{\rm d} N_{l^+l^-} \over {\rm d}\omega {\rm d}^3p} =
C_{em}{\alpha^2_{em}  \over 6 \pi^3} {\rho_V(\omega,\vec{p},T) 
\over (\omega^2-\vec{p}^2) ({\rm e}^{\omega/T} - 1)}
\quad ,
\label{rate}
\end{equation} 
where 
$\alpha_{em}$ is the electromagnetic fine structure constant.
This expression is correct to leading order in the electromagnetic interaction
but contains the full QCD information once the vector spectral function is known.
(see also \cite{Gale:2012xq}).
In the following only the $\vec{p}=0$ case will be discussed.

In \cite{Ding:2010ga} a controlled continuum extrapolation of the vector correlation function
calculated in quenched QCD at $T\simeq 1.45~T_c$ was performed for the first
time.
Those lattice results of \cite{Ding:2010ga} together with the result in the
continuum limit are shown in Fig.~\ref{fig:GVcont}. 

The vector correlation function in the continuum was used to fit to a
phenomenological inspired Ansatz for spectral function, 
\begin{eqnarray}
\rho_{00}(\omega) &=& - 2\pi \chi_q  \omega \delta (\omega)  \ ,
\label{fit00} \\
\rho_{ii} (\omega) &=&  
2\chi_q c_{BW}   \frac{\omega \Gamma/2}{ \omega^2+(\Gamma/2)^2}
+ {3 \over 2 \pi} \left( 1 + k \right) 
\; \omega^2  \;\tanh (\omega/4T)   \ ,
\label{ansatz}
\end{eqnarray}
which contains a Breit-Wigner contribution at small frequencies that gives the
correct $\omega$-dependence in the small frequency limit to determine the
electrical conductivity and a correction parameter $k(T)$ that parameterizes deviations 
from a free spectral function at large energies.
The result of this fit shown in Fig.~\ref{fig:GVcont} demonstrates that 
already this minimal ansatz
provides a good description of current numerical results for the vector
correlation function.

\begin{figure}[thbp]
\centering
\includegraphics[scale=0.643]{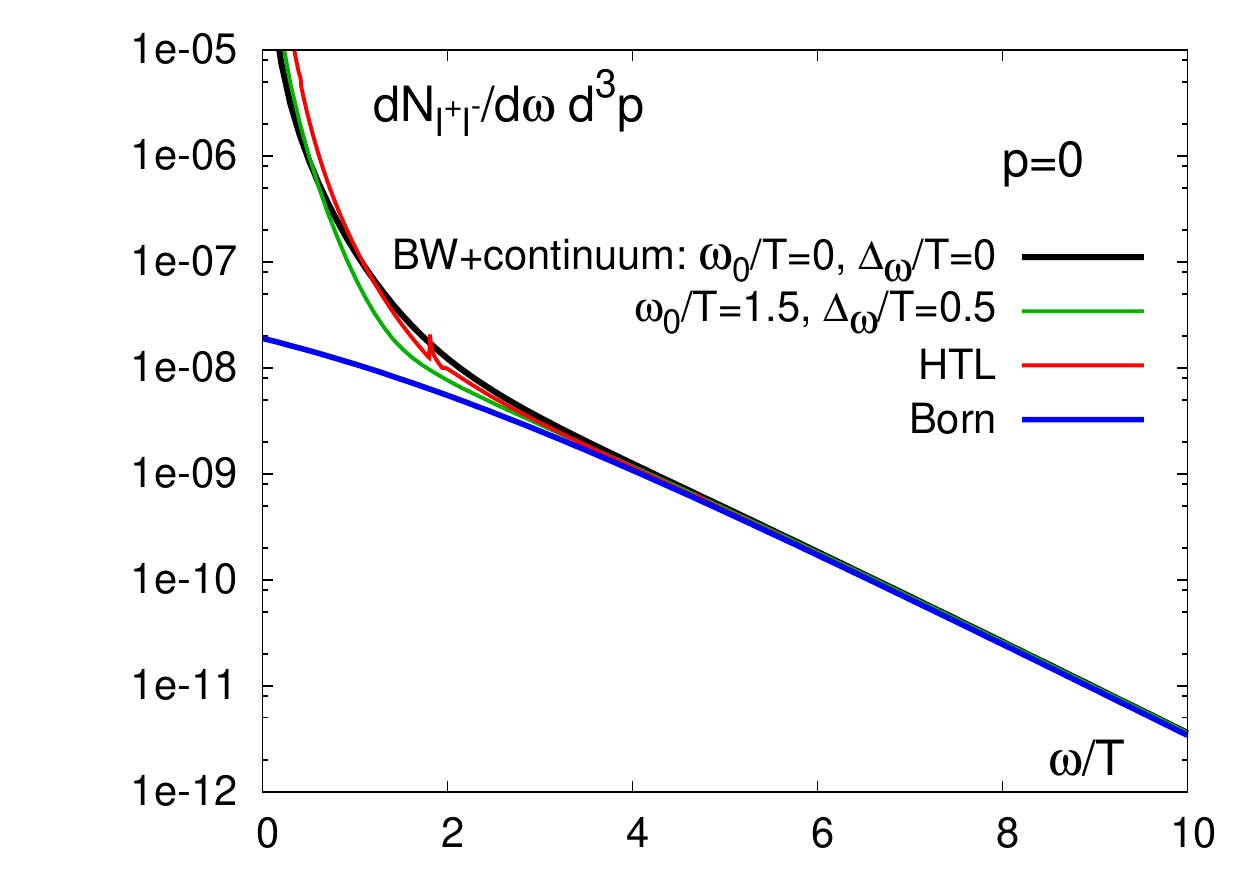}
\includegraphics[scale=0.643]{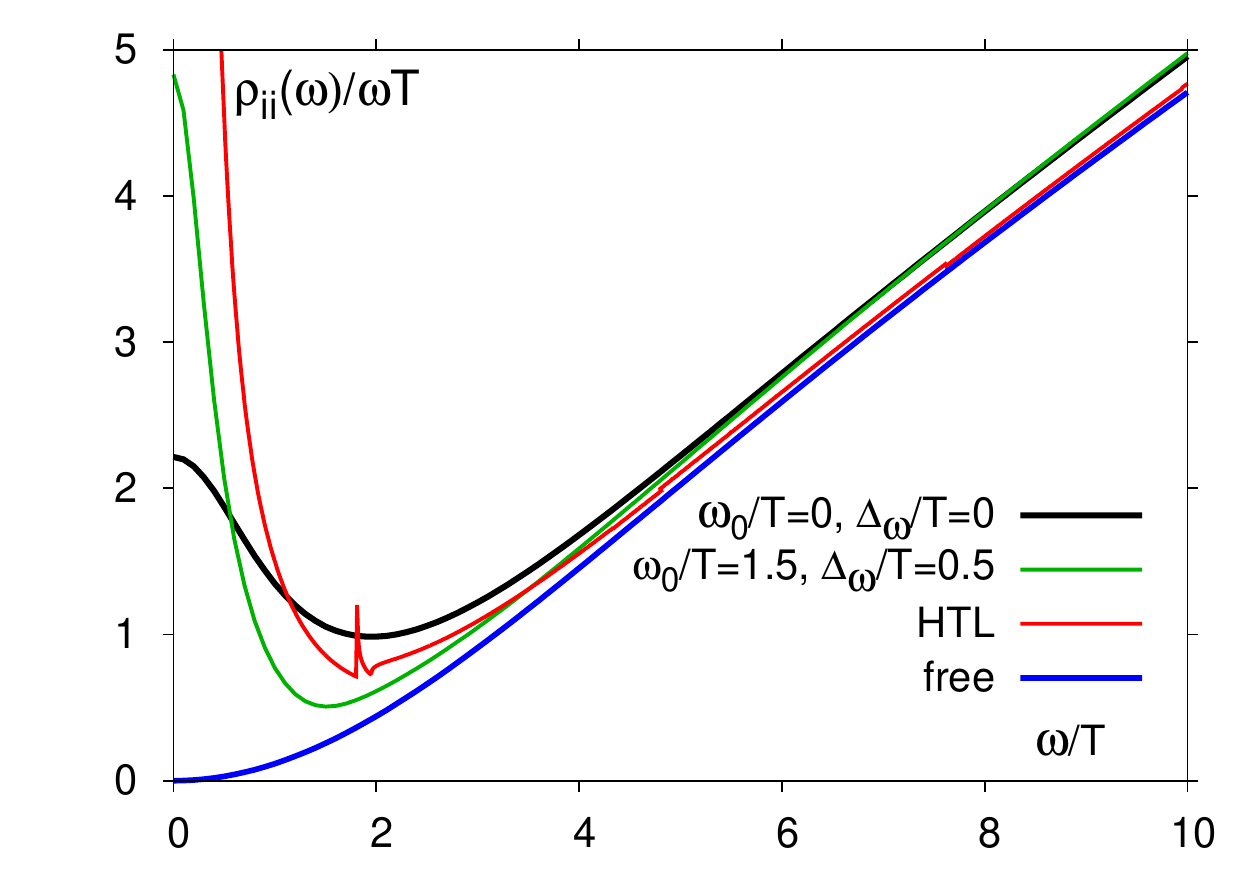}
\caption{
Thermal dilepton rate in 2-flavor QCD (left). 
The HTL curve is for a thermal quark mass $m_T/T=1$ and the 
Born rate is obtained by using the free spectral function. The right hand 
part of the figure shows the spectral functions that entered the calculation 
of the dilepton rate.
(see text and
\cite{Ding:2010ga} for details).}
\label{fig:dilepton}
\end{figure}

Using Eq.~(\ref{rate}) the continuum estimate for the dilepton rate at $T\simeq
1.45~T_c$ in Fig.~\ref{fig:dilepton}(left) are obtained. They merge with the
Born-rate at $\omega ~ \gsim ~ 4$ and are well above this rate for lower energies,
still staying below the HTL result at small $\omega$. In contrast to the HTL result, the limit
$\omega\rightarrow 0$ is well behaved (see Fig.~\ref{fig:dilepton}(right)).
In this
limit the spatial components of the vector spectral function determine the 
electrical conductivity
\begin{equation}
\frac{\sigma}{T} = \frac{C_{em}}{6} \lim_{\omega \rightarrow 0} 
\frac{\rho_{ii}(\omega)}{\omega T} \; .
\label{conduct}
\end{equation}
The vector spectral function 
shown in Fig.~\ref{fig:dilepton} at vanishing energy
leads to an estimate of the electrical conductivity
\begin{equation}
1/3 \ \lsim \ \frac{1}{C_{em}}
\frac{\sigma}{T} \ \lsim \ 1 \quad {\rm at} \quad T\simeq 1.45\ T_c \; .
\label{range}
\end{equation}
Although only a lower and upper limit was estimated, the result demonstrates
that transport coefficients can be reliably extracted from continuum
extrapolated lattice QCD correlation functions which in this respect improves
on earlier results \cite{Gupta:2003zh,Aarts:2007wj}. 
See \cite{Ding:2010ga} for more details and \cite{Burnier:2012ts} for a recent
study using a new technique to extract the spectral function in this limit.

The vector spectral function at light-like 4-momentum 
yields the photon emission rate of a thermal medium,
\begin{equation}
\omega \frac{{\rm d} R_\gamma}{{\rm d}^3p} =C_{em} \frac{\alpha_{em}}{4\pi^2} 
\frac{\rho_{V}(\omega =|\vec{p}|, T)}{{\rm e}^{\omega/T} -1} \ .
\label{photon}
\end{equation}
The emission rate of soft photons, thus can be related to the electrical
conductivity,
\begin{equation}
\lim_{\omega \rightarrow 0} \omega \frac{{\rm d} R_\gamma}{{\rm d}^3p} =
\frac{3}{2\pi^2} \sigma(T) T \alpha_{em} \ .
\label{softphoton}
\end{equation}
Together with Eq.~(\ref{photon}) this yields for the zero energy limit of the thermal 
photon 
rate,
\begin{equation}
\lim_{\omega \rightarrow 0} \omega \frac{{\rm d} R_\gamma}{{\rm d}^3p} =
\left( 0.0004 \ -\ 0.0013 \right) T_c^2 \simeq (1-3)\cdot 10^{-5}\ {\rm GeV}^2
\quad {\rm at} \quad T\simeq 1.45\ T_c  \ .
\label{softphoton2}
\end{equation}
Results of the vector correlation function at non-zero momentum will be
used in the future to estimate $p\neq 0$ dilepton rates and the finite energy
contribution of photons in the QGP.

\section{Charmonium screening masses}
\label{sec:screen}

\begin{figure}[t]
\centering
\includegraphics[scale=0.6]{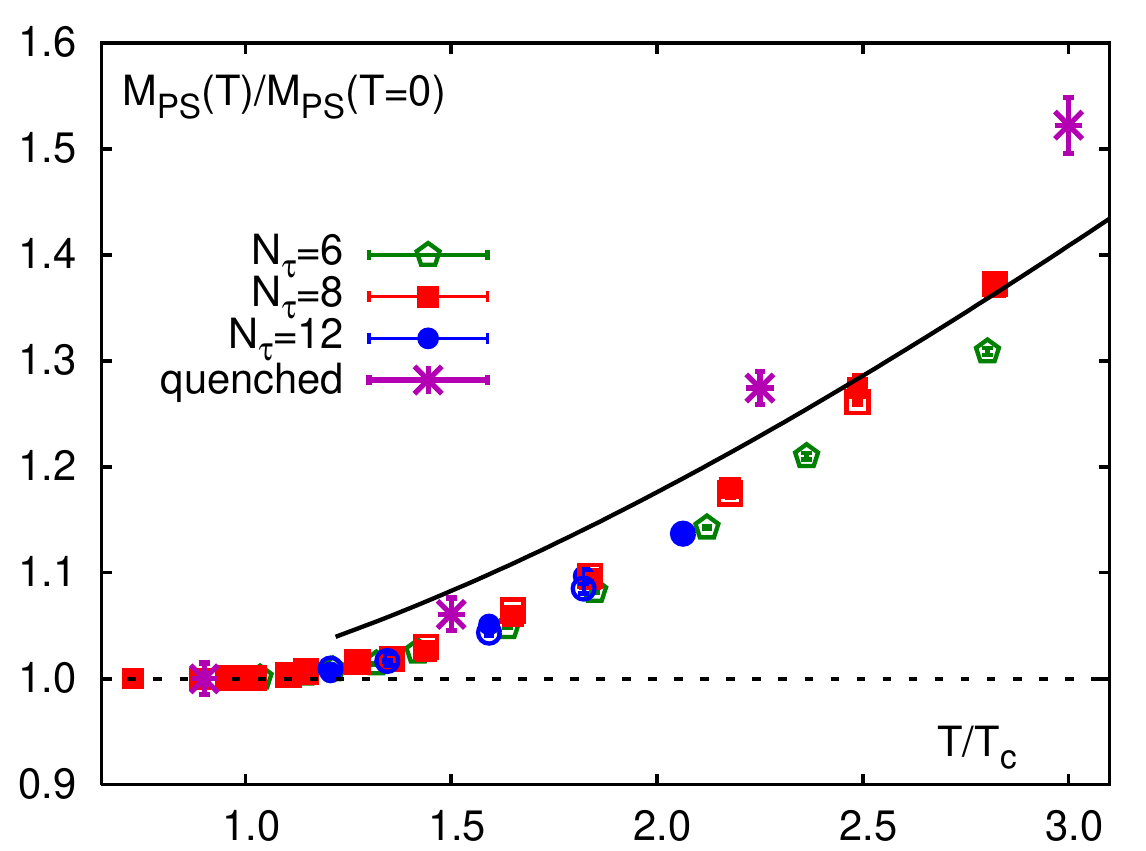}
\includegraphics[scale=0.6]{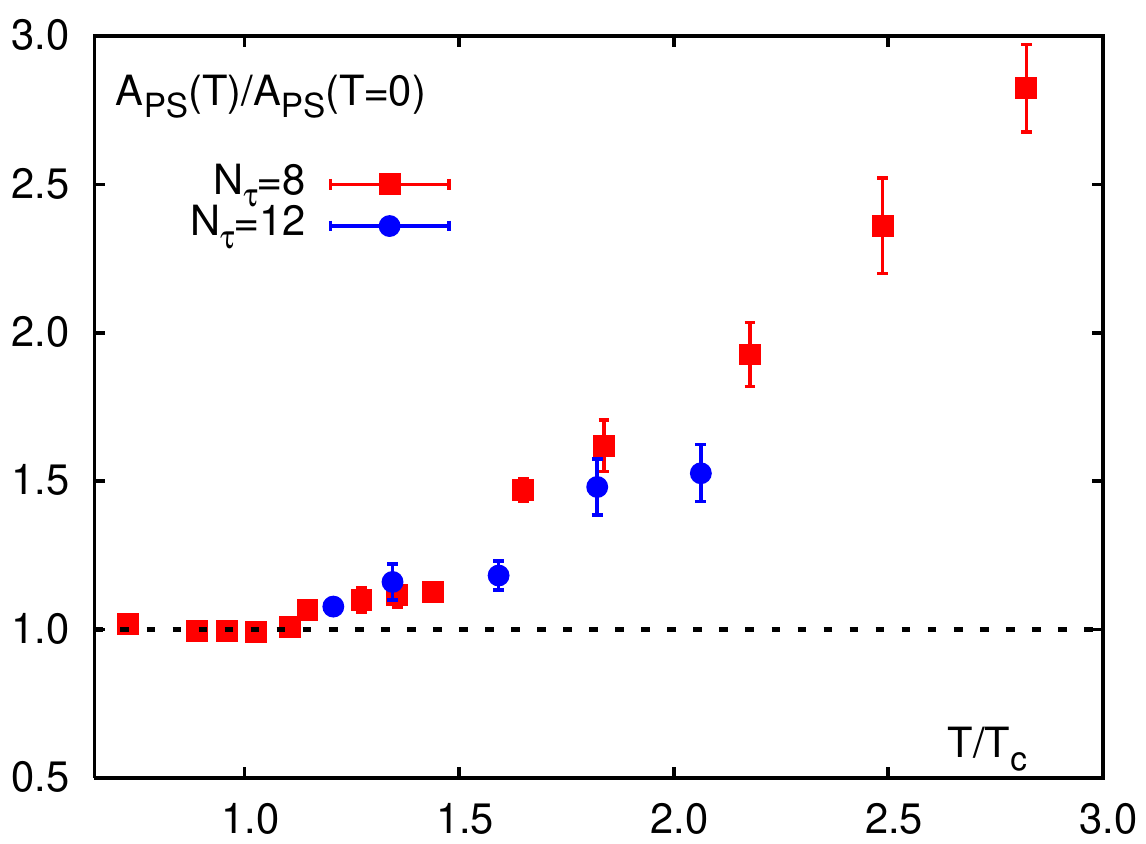}
\caption{Pseudo-scalar screening mass (a) and amplitude (b) divided
by the corresponding zero temperature values as function of the temperature.
The solid black line is the free quark result.
The plots are from \cite{Karsch:2012na}.}
\label{fig:m_ps}
\end{figure}

High precision data at a large number of
Euclidean time separations are needed to extract spectral properties from
temporal correlation function. An alternative approach that may provide
additional information on the in-medium modifications of spectral functions and
medium effects of mesonic states are spatial correlation functions,
\begin{equation}
G(z,T)=\int dx dy \int_0^{1/T} d \tau \langle J(x,y,z,\tau) J(0,0,0,0) \rangle,
\label{eq:Gz}
\end{equation}
which are related to the momentum dependent spectral function through
\begin{equation}
G(z,T)=\int_{0}^{\infty} \frac{d \omega}{2\pi}\frac{2}{\omega} \int_{-\infty}^{\infty} \frac{d p_z}{2\pi} e^{i p_z z} \rho(\omega,p_z,T).
\end{equation}
At large distance, $zT>1$, where medium effects are expected to be largest,
$G(z,T)$ decays exponentially allowing to define a screening mass $M_{scr}$. If
bound states appear in the spectral function, the screening mass is determined
by the mass $M$ of the lowest lying meson state. On the other hand for
unbound quark antiquark pairs at high temperature, $M_{scr}$ is given by the
lowest non-vanishing Matsubara frequency $2 \sqrt{(\pi T)^2+m_c^2}$ where $m_c$
is the quark mass. Those two limits and the transition between them may serve
a an indication for significant modifications and dissociation of heavy quark
meson states.

In Fig.~\ref{fig:m_ps} the results for the charmonium screening masses calculated
for a dynamical 2+1 flavor lattice QCD from \cite{Karsch:2012na}
are shown, normalized to the zero temperature value. The change in the
temperature dependence becomes apparent around $1.5~T_c$ where deviations from
the zero temperature limit become large and the behavior becomes compatible
with that of unbound $c\bar c$ pairs. This is an indication that charmonium
states melt around this temperature which is a consistent picture with
the findings from the study of temporal charmonium correlators in quenched QCD 
\cite{Ding:2012sp} that will be discussed in the following section.


%
\newcommand{\rmi}[1]{{\! \mbox{\scriptsize #1}}}
\newcommand{\fr}[2]{{\frac{#1}{#2}}}
\newcommand{\re}  {\mathop{\rm Re}}
\newcommand{\Tr}  {\mathop{\rm Tr}}

\section{Charmonium spectral functions and heavy quark diffusion}
\label{sec:charm}

\begin{figure}[t]
\centering
\includegraphics[scale=0.085]{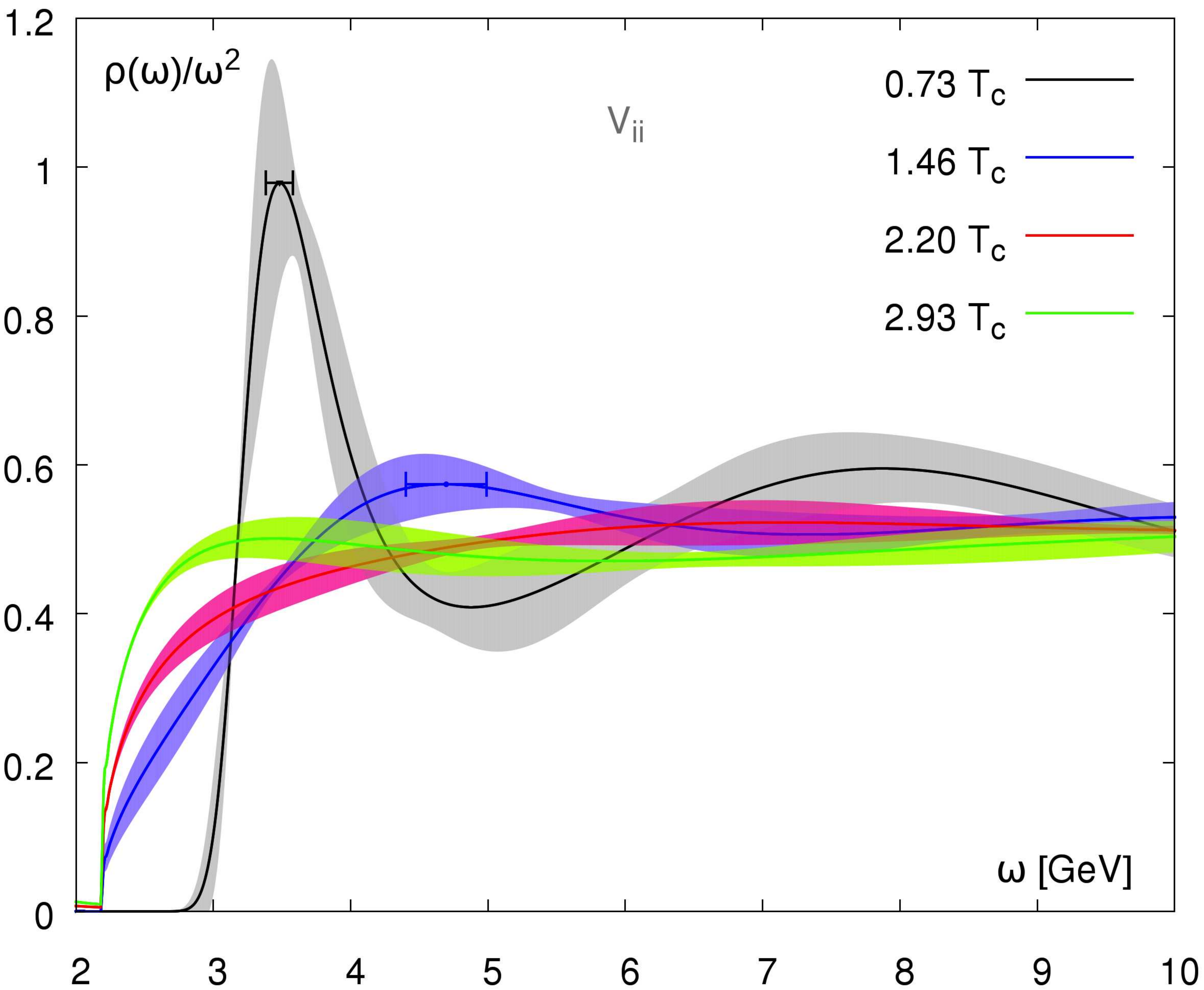}
\includegraphics[scale=0.085]{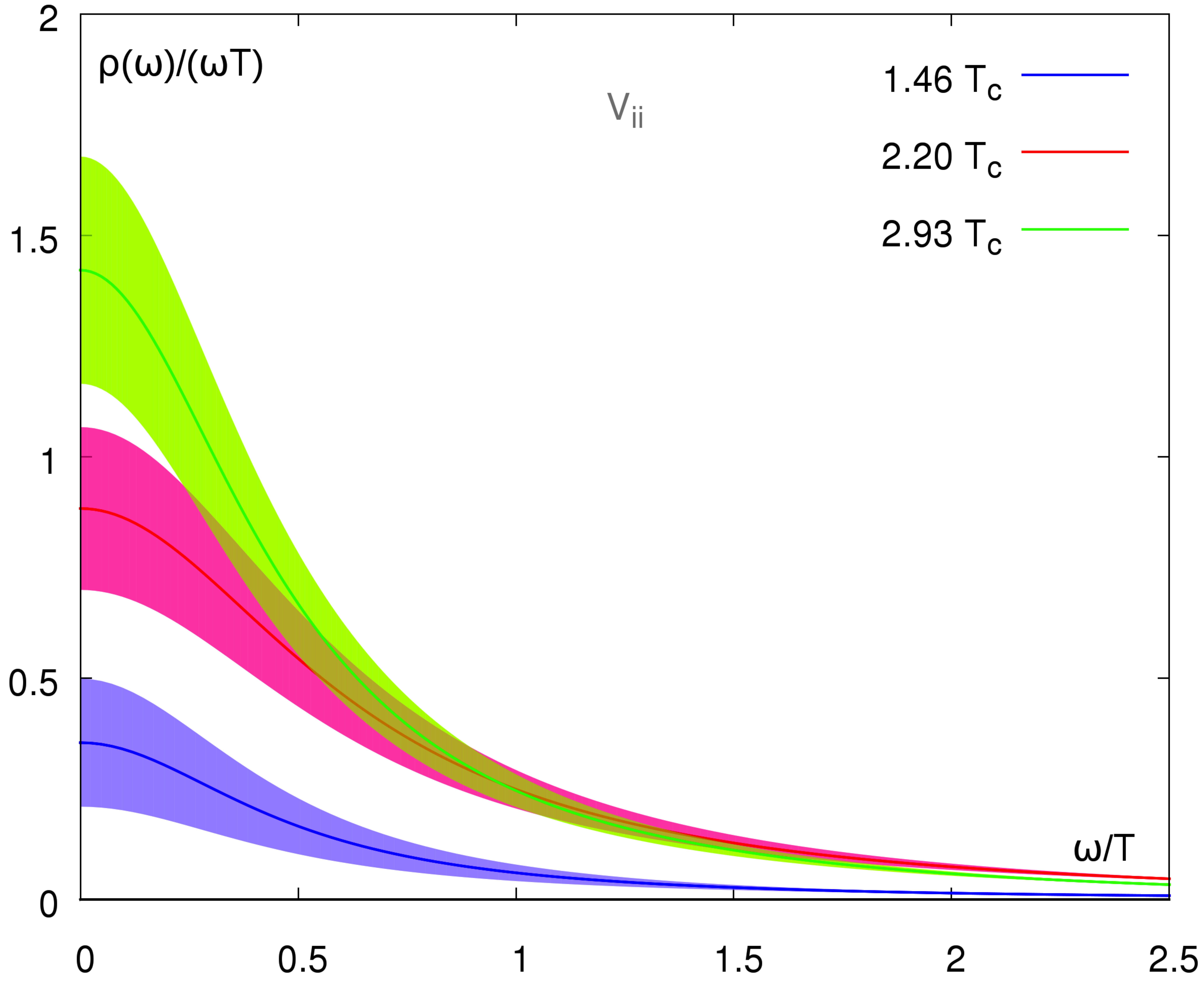}
\caption{
Results for the charmonium spectral functions in the vector
channel. Shaded areas are statistical errors from Jackknife analyses. 
Horizontal error bars at the first peaks of spectral functions at $0.73~T_c$ an
$1.4~T_c$ 
indicate the statistical uncertainties of the peak location obtained from
the Jackknife analyses.
(see
\cite{Ding:2012sp} for details).}
\label{fig:charm}
\end{figure}

The properties of charmonium states at finite temperature were studied in 
\cite{Ding:2012sp} in quenched lattice QCD on large and fine isotropic lattices.
Although no continuum extrapolation was performed, the results for lattice
sizes of $128^3\times N_\tau$ for $N_\tau$ between 16 and 96 and lattice
spacing $a$ down to $0.01~{\rm fm}$ are already close to the continuum and
cut-off effects in the spectral function are well separated from the physically
interesting energy regions, as $a m_c$ is much smaller than one.
In \cite{Ding:2012sp} a detailed analysis of statistical and systematic
uncertainties and default
model dependencies in the maximum entropy method (MEM) used to determine the
spectral functions from the lattice calculations of the charmonium correlation
functions was performed.

The results of this study, shown in Fig.~\ref{fig:charm}
for the vector channel,
suggest that charmonium states are dissociated already at $1.46~T_c$.
While at $0.73~T_c$ stable and reliable ground state peaks are found, at the
higher temperatures in this study no signals for bound state contributions 
but rather a threshold enhancement are seen.

The slope of the vector spectral function in the $\omega\rightarrow 0$ limit
again determines a transport coefficient,
the heavy quark diffusion constant $D$,
\begin{eqnarray}
D = \frac{1}{6\chi^{00}}\lim_{\omega\rightarrow0}\sum_{i=1}^{3}\frac{\rho^{V}_{ii}(\omega,\vec{p}=0,T)}{\omega} ,
\label{eq:HQ_diffusion_formula}
\end{eqnarray}
where $\chi^{00}$ is the quark number susceptibility.
While compatible with zero below $T_c$, almost temperature independent values
of $2\pi T D$ around two at $T=1.46, 2.20$ and $2.93$
were obtained in \cite{Ding:2012sp} (see Fig.~\ref{fig:DT}).
 
This is a further indication of strong medium effects and 
$J/\Psi$ being melted at $1.46~T_c$ and higher temperatures. Together with the
results for the charmonium screening masses of the previous section this gives
a consistent picture and shows that the interesting temperature region for
medium effects and the study for dissociation of charmonium is between $T_c$
and $1.5~T_c$.
Future studies at temperatures closer to $T_c$ using continuum extrapolated
correlation functions may allow a more detailed determination of the dissociation
patterns of the different charmonium states in the quark gluon plasma.

\section{Bottomonium from lattice NRQCD}
\label{sec:nrqcd}

\begin{figure}[t]
\centering
\includegraphics[scale=0.6]{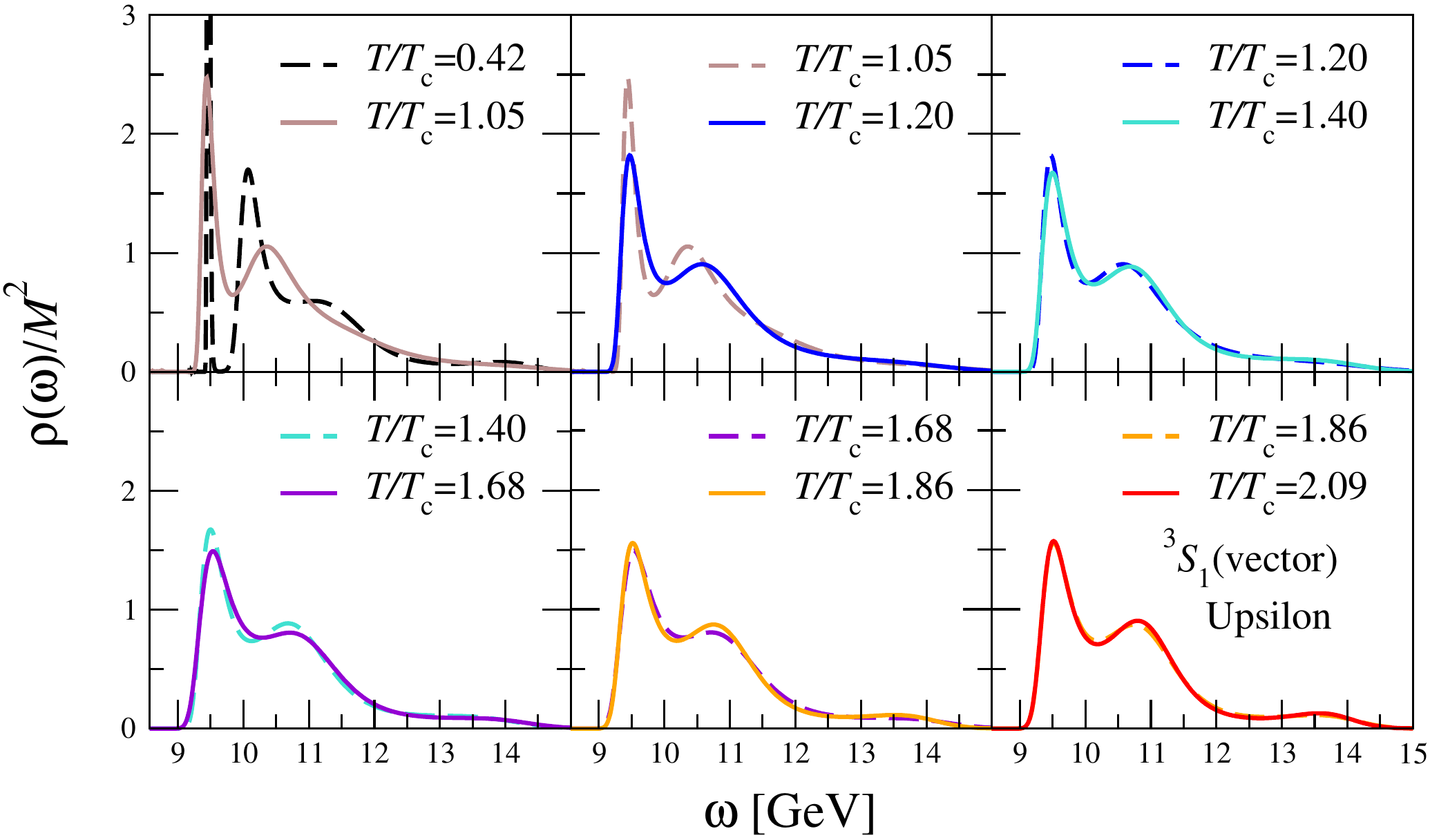}
\caption{
Spectral functions $\rho(\omega)$ in the vector channel ($\Upsilon$),
normalized with the heavy quark mass. This plot is from \cite{Aarts:2011sm}.
}
\label{fig:upsilon}
\end{figure}

At the energies probed at the
LHC bottomonium spectroscopy becomes an important tool for the analysis of
the produced hot medium. It may offer a cleaner probe compared to charmonium as
feed-down effects from heavier quarkonium states are missing and regeneration
from the thermal medium will be negligible. Furthermore the dissociation
temperatures are far higher than for charmonium and therefore well separated
from the hadronization stage.
First results from the CMS experiment at LHC show a suppression of the
higher excited states $\Upsilon(2S+3S)$ \cite{Reed:2011fr} and already indicate the
importance of bottomonium in this respect.
Therefore a detailed knowledge on the
behavior and dissociation temperatures of bottomonium states is
of fundamental importance.

For bottom quarks the heavy quark mass scale can be integrated out in terms of an
effective field theory, nonrelativistic QCD (NRQCD). In a lattice
discretization of this effective theory where the physics above the scale $m_b$
is encoded in the perturbatively determined coefficients of the NRQCD
Lagrangian, the spatial lattice spacing acts as a short distance cut-off and
must satisfy $m_b a_s ~ \gsim ~ 1$ \cite{Aarts:2011sm,Aarts:2010ek,Fingberg:1997qd}. 
The drawback of this is that no continuum limit
is possible in lattice NRQCD and furthermore that cut-off effects are present
in the spectral function already in the energy region relevant for bottomonium
excited states as the cut-off effects and also the accessible energy region is
determined by the spatial lattice spacing.

In contrast to a computationally expensive inversion of the fermion matrix,
NRQCD propagators are calculated as an initial value problem. 
Information on transport properties are exponentially suppressed which is 
a further consequence of the quark mass scale acting as a low energy cut-off
at $2 m_b$. Nevertheless this facilitates the extraction of spectral properties in
the bound state region as no constant contribution appear in the NRQCD
correlation functions.

Despite the limitations of the method, the results of \cite{Aarts:2011sm}
demonstrate the benefits of lattice NRQCD. Their results for the spectral
function in the vector channel ($\Upsilon$) are shown in
Fig.~\ref{fig:upsilon}. It suggests that the ground state survives up to the
highest temperature analyzed in this work and indicates the melting of higher
excited states at temperatures above $1.4~T_c$. 
See \cite{Aarts:2011sm} for a detailed discussion of the results.

Especially for the behavior of the excited states a detailed analysis of the
systematic uncertainties of this approach is still needed,
especially to understand the cut-off effects and default model dependencies of
the maximum entropy method used to extract the spectral functions.  
In addition lattice NRQCD results will be useful for potential model studies of quarkonium
systems and will give additional input for future full relativistic lattice
QCD calculations of bottomonium states.

\section{Heavy quark momentum diffusion coefficient}
\label{sec:momdiff}

\begin{figure}[t]
\centering
\includegraphics[scale=0.31]{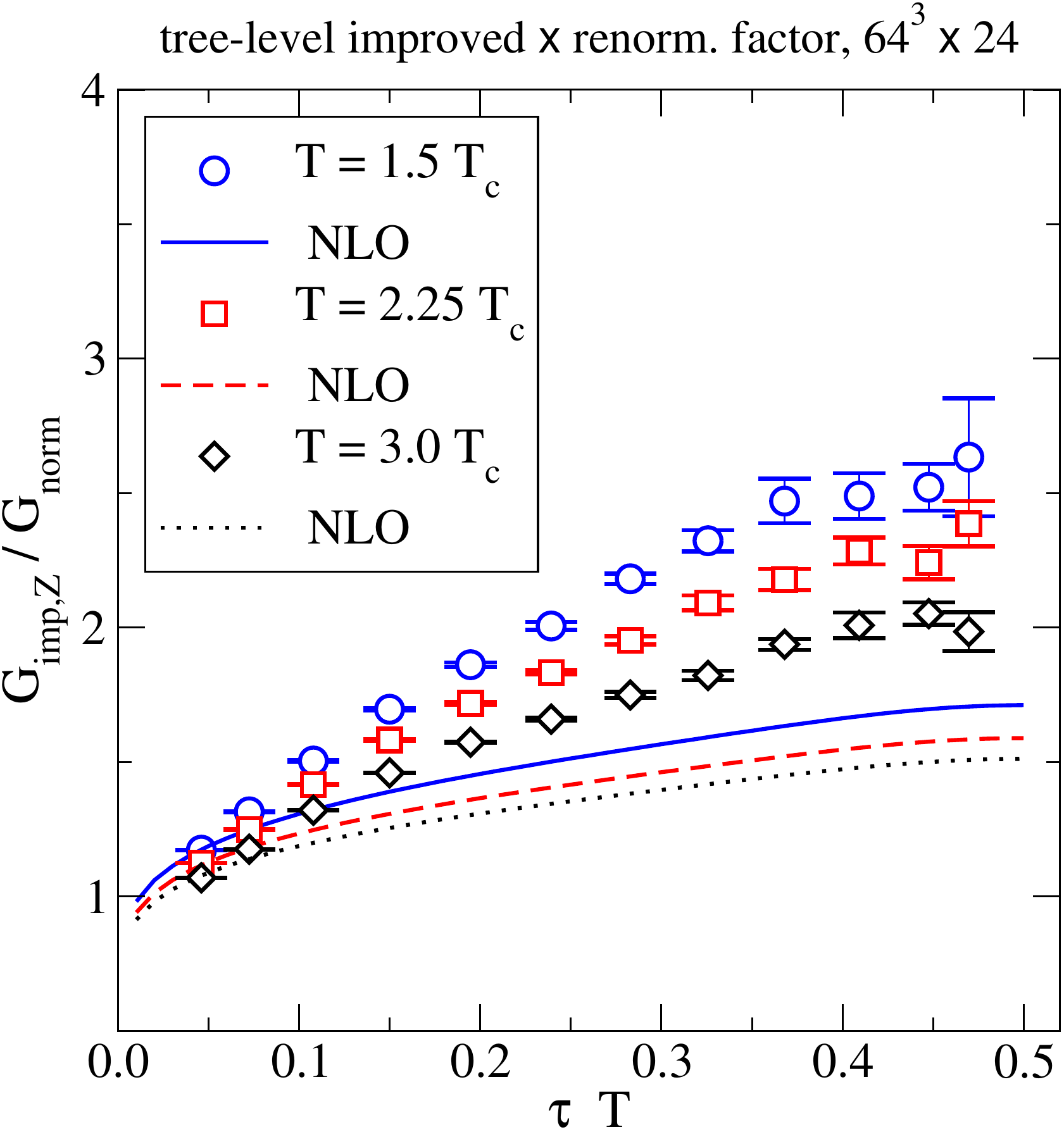}
\includegraphics[scale=0.31]{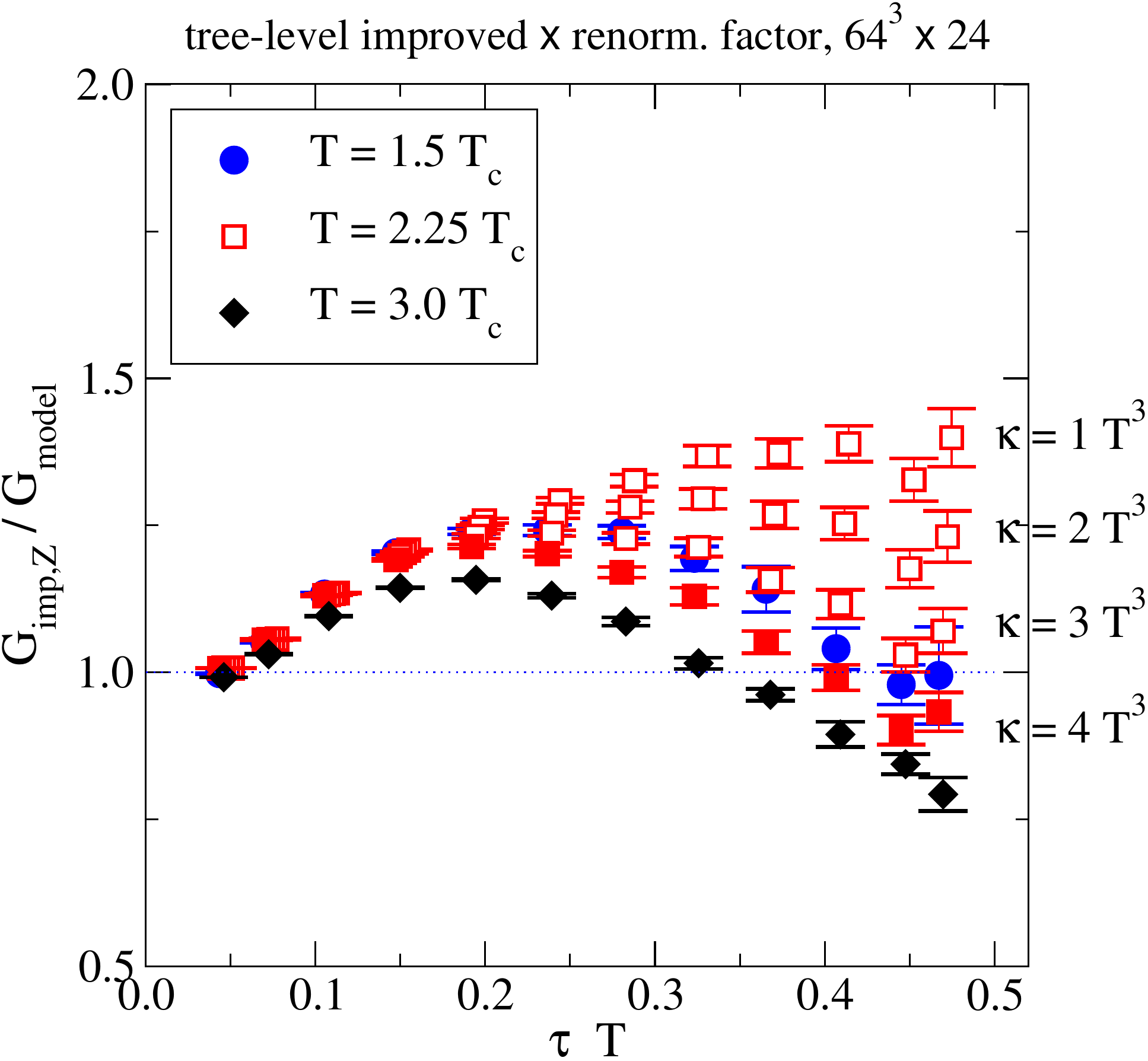}
\caption{
 Left: Comparison of lattice with the NLO weak-coupling
 expansion~\cite{Burnier:2010rp}.
 Right: Like on the left but normalized to the model of
 eq.~\ref{model}. Closed symbols correspond to $\kappa = 4 T^3$,
 open ones to $T = 2.25T_{\rm c}$ (see
\cite{Francis:2011gc} for details).
}
\label{laine}
\end{figure}

Effective field theory methods are also important to derive operators to
extract transport coefficients from lattice QCD. In the large quark mass limit
Heavy Quark Effective Theory (HQET) can be used to derive a purely gluonic
operator which is related to the ''momentum diffusion coefficient'' in the
low-frequency limit,
\begin{eqnarray}
 \kappa = \lim_{\omega\to 0} \frac{2 T \rho_\rmi{\,E}(\omega)}{\omega}
 \;. \label{intercept} 
\end{eqnarray}
Heavy quarks carry a color charge and are therefore subject to a colored
Lorentz force. Like with other transport coefficients the corresponding ``low-energy
constants'' are easiest to define at vanishing three-momentum when
the Lorentz-force is proportional to the electric field strength. 
This leads to 
a ``color-electric correlator''~\cite{CasalderreySolana:2006rq,CaronHuot:2009uh},  
\begin{eqnarray}
 G_\rmi{\,E}(\tau) \equiv - \fr13 \sum_{i=1}^3 
 \frac{
  \Bigl\langle
   \re\Tr \Bigl[
      U(\fr{1}{T};\tau) \, gE_i(\tau,\vec{0}) \, U(\tau;0) \, gE_i(0,\vec{0})
   \Bigr] 
  \Bigr\rangle
 }{
 \Bigl\langle
   \re\Tr [U(\fr{1}{T};0)] 
 \Bigr\rangle
 }
 \;, \label{GE_final}
\end{eqnarray}
where $gE_i$ denotes the  
color-electric field, $T$ the temperature, and $U(\tau_2;\tau_1)$
a Wilson line in Euclidean time direction. 
If the corresponding spectral function, $\rho_\rmi{\,E}$,  
can be extracted~\cite{Meyer:2011gj}, then the ``momentum 
diffusion coefficient'', often denoted by $\kappa$, can be 
obtained from Eq.~(\ref{intercept}).
According to non-relativistic linear response 
relations (valid 
for $M \gg \pi T$, where $M$ stands for a heavy quark pole mass)
the corresponding ``diffusion coefficient'' 
is given by $D = 2 T^2/\kappa$. 

Due to the gluonic nature of the operator Eq.~(\ref{GE_final}) large 
fluctuations lead to weak signal/noise ratio. Using clever improvement
algorithms in addition with tree-level improvement to reduce cut-off effects
and a perturbative renormalization of the operator leads to results that show
almost no lattice artifacts.\\
In Fig.~\ref{laine}(left) the results of \cite{Francis:2011gc} are shown in
comparison with a NLO weak-coupling expansion, both normalized by 
the leading order \cite{CaronHuot:2009uh} correlation function.
To extract the diffusion coefficient we use a model for the spectral function 
\begin{eqnarray}
 \rho_\rmi{\,model}(\omega)
 \equiv \mathop{\mbox{max}}
 \Bigl\{ \rho_\rmi{\,NLO}(\omega) , \frac{\omega \kappa}{2 T} \Bigr\}
 \;, \label{model}
\end{eqnarray}
with the free parameter $\kappa$ representing directly the
momentum diffusion coefficient according to Eq.~(\ref{intercept}),
and the corresponding Euclidean correlator computed from
\begin{eqnarray}
 G_\rmi{\,model}(\tau) \equiv
 \int_0^\infty
 \frac{{\rm d}\omega}{\pi} \rho_\rmi{\,model}(\omega)
 \frac{\cosh \left( \left(\frac{1}{2} - \tau T \right) \frac{\omega}{T} \right) }
 {\sinh\left(\frac{\omega}{2 T}\right)}
 \;. \label{int_rel}
\end{eqnarray}
Results for the ratio of the lattice data and the model correlation function
for different values of $\kappa$ are shown in Fig.~\ref{laine} and allow for an
estimate for the momentum diffusion constant at the three temperatures
analyzed. Converting those estimates to the usual diffusion coefficient $D=2
T^2/\kappa$ one obtains $DT\sim (0.5\cdots 0.8)$ which is slightly larger than
the values obtained for the charm case \cite{Ding:2012sp} (see section
\ref{sec:charm}) and on this qualitative level consistent with a similar
analysis in \cite{Banerjee:2011ra}.

Although the values for $\kappa$ and $DT$ are to be understood as a first rough
estimate, it indicates that a consistent picture appears to emerge. Improving
the lattice data for the correlation function and performing the continuum
limit, which is possible in this approach, will allow to test improved models
for the spectral function and use methods like the maximum entropy method or
other sophisticated methods to perform the analytic continuation like the one proposed
and tested in \cite{Burnier:2011jq,Burnier:2012ts}.

\begin{figure}[t]
\centering
\includegraphics[scale=0.8]{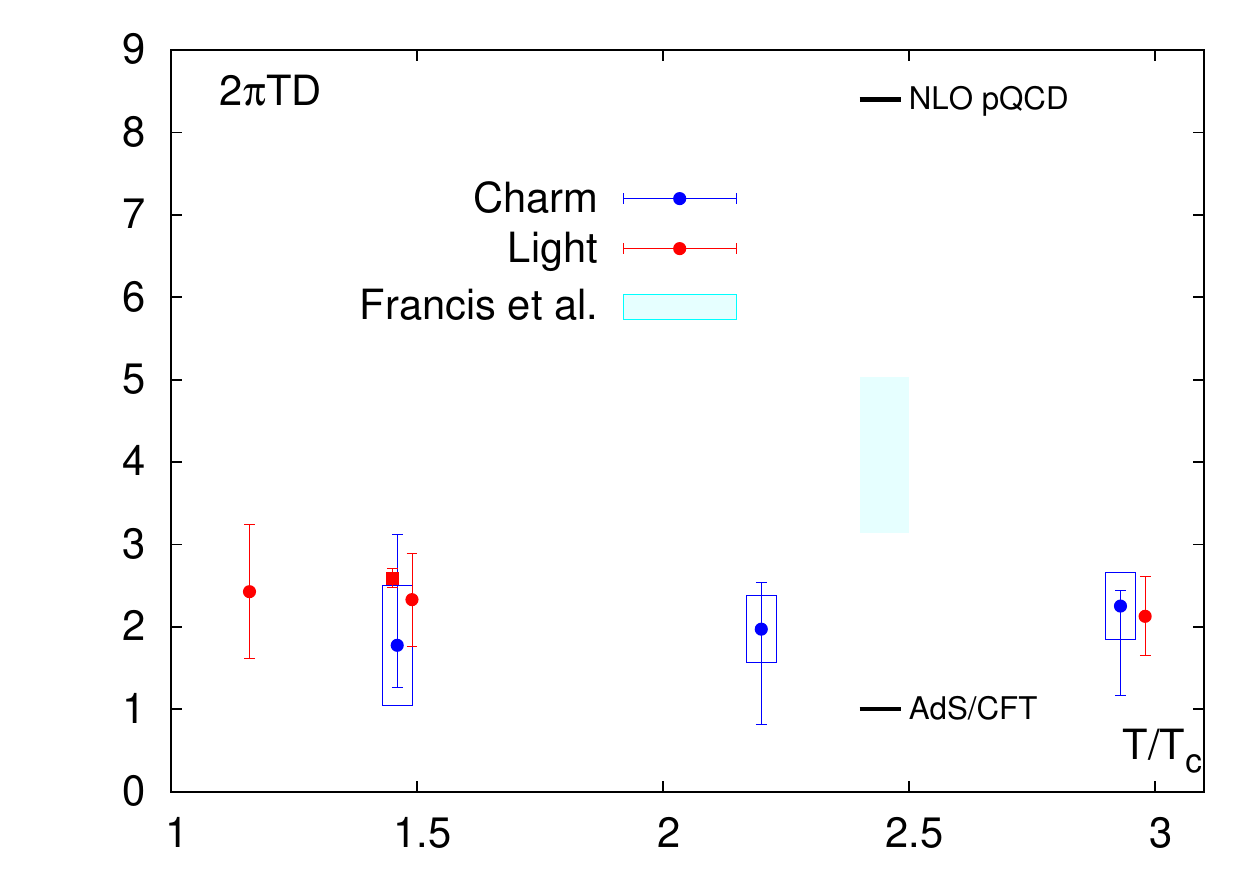}
\caption{
Comparison of various estimates for the heavy quark diffusion constant (see
the text for more details).
}
\label{fig:DT}
\end{figure}

\section{Conclusions and outlook}
Lattice QCD calculations have proven to provide important information on the
in-medium modifications of hadronic states for light up to bottom quarks. High
quality results and sophisticated methods to determine spectral properties
from lattice correlation functions allow a reliable estimate of transport
coefficients.
This is accompanied by effective field theory methods used in the heavy quark
mass limit. Especially for the analysis of bottomonium and even more for the
determination of transport coefficients from lattice correlation functions 
additional information can be gained from this combination.

Although most results are still obtained in the quenched limit or for rather
large light quark masses, a consistent picture of medium modifications and the
dissociation of quarkonia as well as of transport coefficients needed to
describe the thermal medium and the behavior of heavy quarks in this medium
emerges. 

For future charmonium studies the interesting temperature region between $T_c$
and $1.5~T_c$ shall be explored in the future as the recent results indicate that charmonium
is melted already above those temperatures. At least for the bottomonium ground
states higher temperatures seem to be relevant and full relativistic lattice
calculation are necessary to confirm the findings of the exploratory NRQCD
study in this sector.

In Fig.~\ref{fig:DT} we compare the values for the diffusion coefficients
obtained from the analysis of the charmonium spectral function in quenched QCD
\cite{Ding:2012sp} and the heavy quark momentum diffusion coefficient in the large quark
mass limit \cite{Francis:2011gc} (light blue band). 
On this qualitative level all results are
compatible with estimates derived from the
heavy flavor $R_{AA}$ and $v_w$ PHENIX
data \cite{Adare:2010de,Adare:2006nq}. See also \cite{Riek:2010fk} for a
determination of the diffusion constant using heavy quark free energies in a
T-matrix approach 
and \cite{He:2012df} for a recent
discussion of heavy quark diffusion constants.
Just for comparison we also
show the result obtained from the electrical conductivity in the light quark
sector \cite{Ding:2010ga,Francis:2011bt}.

The results of section \ref{sec:nrqcd} and \ref{sec:momdiff} show the
opportunity and applicability of effective field theory methods and operators
derived from it for the study of heavy quark spectral and transport properties
in lattice QCD calculations. Despite their limitations they will allow to
constrain models, e.g. for potential model calculations
\cite{Rothkopf:2012et,Burnier:2012ib},
and contribute additional information and input needed in full
relativistic lattice QCD calculations in the future especially for the
difficult task of extracting spectral properties from hadronic correlation
functions.

In a similar way as for the light quark case discussed in section
\ref{sec:dilepton}, an extrapolation of the correlation functions to the
continuum will be required in the heavy quark sector to remove any lattice
discretization effects.
Although at the moment only possible in quenched QCD this will allow a more
quantitative determination of spectral and transport properties for quarkonium
systems.
This will lay the foundation for the analysis of quarkonium correlation
functions in full QCD with realistic light quark mass becoming available in the
coming years at least for moderate temporal lattice extensions and will allow
to determine the effects of light quark masses on spectral properties of heavy
mesons.

\bibliographystyle{elsarticle-num}
\bibliography{hp12}

\end{document}